# StaPep: an open-source tool for the structure prediction and feature extraction of hydrocarbon-stapled peptides

Zhe Wang[1,2], Jianping Wu[1,3], Mengjun Zheng[4], Chenchen Geng[4], Borui Zhen[4], Wei Zhang[1,2], Hui Wu[5], Zhengyang Xu[6], Gang Xu[1,]*, Si Chen[7,]*, Xiang Li[4,]*

[1]Institute of Bioengineering, College of Chemical and Biological Engineering, Zhejiang University, Hangzhou, China

[2]Hangzhou VicrobX Biotech Co. , Ltd., China

[3]ZJU-Hangzhou Global Scientific and Technological Innovation Center, Hangzhou, China

[4]School of Pharmacy, Second Military Medical University, Shanghai, China

[5]Huadong Medicine Co., Ltd., China

[6]School of Chemical and Environmental Engineering, Shanghai Institute of Technology, Shanghai, 201418, PR China

[7]School of Medicine, Shanghai University, Shanghai, China

*Please address correspondence to Dr. Xiang Li at xiangli@smmu.edu.cn, Dr. Si Chen at caroline-sisi-chen@hotmail.com and Dr. Gang Xu at xugang_1030@zju.edu.cn

**Abstract**

Many tools exist for extracting structural and physiochemical descriptors from linear peptides to predict their properties, but similar tools for hydrocarbon-stapled peptides are lacking. Here, we present StaPep, a Python-based toolkit designed for generating 2D/3D structures and calculating 21 distinct features for hydrocarbon-stapled peptides. The current version supports hydrocarbon-stapled peptides containing 2 non-standard amino acids (norleucine and 2-aminoisobutyric acid) and 6 nonnatural anchoring residues (S3, S5, S8, R3, R5 and R8). Then we established a hand-curated dataset of 201 hydrocarbon-stapled peptides and 384 linear peptides with sequence information and experimental membrane permeability, to showcase StaPep's application in artificial intelligence projects. A machine learning-based predictor utilizing above calculated features was developed with AUC of 0.85, for identifying cell-penetrating hydrocarbon-stapled peptides. StaPep's pipeline spans data retrieval, cleaning, structure generation, molecular feature calculation, and machine learning model construction for hydrocarbon-stapled peptides. The source codes and dataset are freely available on Github: https://github.com/dahuilangda/stapep_package.

## 1 Introduction

Peptides are an important class of molecules with promising applications in the field of drug discovery. They exhibit high specificity and affinity towards their targets, as well as reduced toxicity compared to small molecule drugs.[1] Despite these advantages, the clinical translation of peptides has been limited due to their poor cell permeability, proteolytic susceptibility, and rapid clearance from the body.[2] The all-hydrocarbon peptide stapling strategy have been developed to improve the proteolytic stability and membrane permeability of linear peptides.[3]

Despite the remarkable promise of this strategy for improving the capacity of peptides for cellular uptake, the specific criteria for designing cell-penetrating stapled peptides remain unclear. Currently, design strategies are largely based on trial-and-error or cumulative empirical observations.[4] Studies have reported that several features of stapled peptides, including the positioning of the staple at the amphipathic boundary, hydrophobic related indexes, helical content, charge, peptide sequence, composition and placement of the staple, and interaction with membranes, correlated with their capacity for cell penetration[4,5]. However, until now, there is no model available to distinguish between cell-penetrating hydrocarbon-stapled peptides and non-membrane permeable molecules, which hinders the wider application of hydrocarbon-stapled peptides for cellular and in vivo analyses.

Currently, molecular features serve as a fundamental element in the application of artificial intelligence models to cheminformatics and drug discovery. Many tools have been developed to calculate molecular features, including physicochemical properties and indices, for linear peptide amino-acid sequences. These tools include IFeature[6], Peptides[7], PDAUG[8], Pepfun[9], PyBioMed[10], OpenBabel[11], CycloPs[12],

ExPASy-ProtParam[13], AAindex[14], modlAMP[15], PyDPI[16], PROFEAT[17], protr/ProtrWeb[18], Rcpi[19], EMBOSS Pepstats[20], BioPerl[21], CAMP[22], PseAAC[23] and so on. Based on the calculated peptide properties, many web-based tools have been developed to identify cell-penetrating peptides, including TargetCPP[24], StackCPPred[25] and so on[26]. However, there is a lack of tools for calculating structural and physiochemical descriptors of hydrocarbon-stapled peptides, which limits the development of AI models for predicting their cellular permeability.

To address these limitations, we developed a comprehensive and extensible toolkit (StaPep) that integrated diverse functionalities including data acquisition, data cleaning, generation of 2D/3D structures, computation of molecular features and construction of machine learning (ML) models for hydrocarbon-stapled peptides. A self-built data set, extracted from 123 articles and comprising 201 hydrocarbon-stapled peptides and 384 linear peptides with sequence information and experimental membrane permeability data, was established. This dataset was applied as an example to assist users in using StaPep for artificial intelligence projects. An ML model, utilizing features calculated by StaPep, was developed to identify cell-penetrating hydrocarbon-stapled peptides, achieving an AUC of 0.85. This study offers valuable insights for the future strategic design of cell-penetrating hydrocarbon-stapled peptides with novel applications, thereby accelerating the development of such peptides for targeted biological and therapeutic uses. Additionally, StaPep can also facilitate other properties predictions of hydrocarbon-stapled peptides, such as pharmacokinetic and pharmacodynamic properties.

## 2 Results

In this study, we have developed StaPep, a Python-based open-source toolkit designed to extract 21 kinds of sequential and structural features from hydrocarbon-stapled peptide (Figure 1). StaPep empowers users to build benchmark datasets, apply various filters and develop ML models using these extracted features. To the best of our knowledge, this is the first toolkit dedicated to calculating molecular features of hydrocarbon-stapled peptide and constructing ML models aimed at distinguishing between cell-penetrating peptides and non-membrane-permeable peptides.

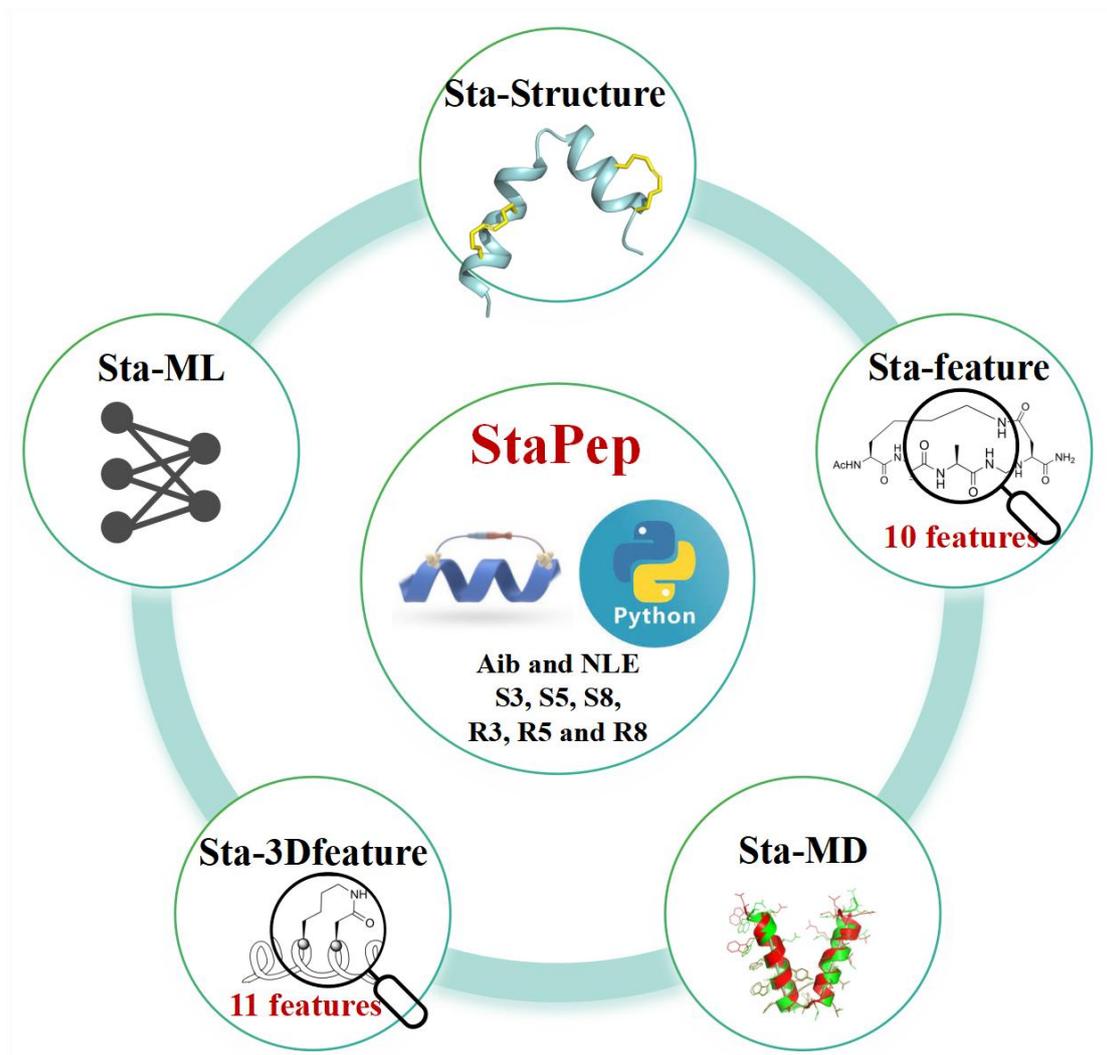

**Figure 1** The main modules of the StaPep.

## 2.1 Example dataset analysis

In our study, we showcased StaPep's capabilities through a case study where we developed a LightGBM model to predict the cell-penetrating potential of hydrocarbon-stapled peptides. A high-quality dataset consisting of 325 cell-penetrating peptides and 260 non-cell penetrating peptides (Figure 2A, left), accompanied by sequence information and experimental data on membrane permeability, was manually collected from references. The dataset was then refined, removing redundancies and improving data quality, resulting in a revised count of 288 cell-penetrating and 218 non-cell-penetrating peptides (Figure 2B, left). It's worth noting that in our hydrocarbon stapled-peptide dataset, the vast majority of samples, exceeding 80%, are positive (permeable). And relatively few, less than 20%, are negative (impermeable), as depicted in Figure 2A and 2B (right). To balance the dataset, we augmented it by including linear peptides with sequence information and experimental membrane permeability data. This augmentation effectively equalized the dataset, achieving a nearly 1:1 ratio between permeable and impermeable peptides.

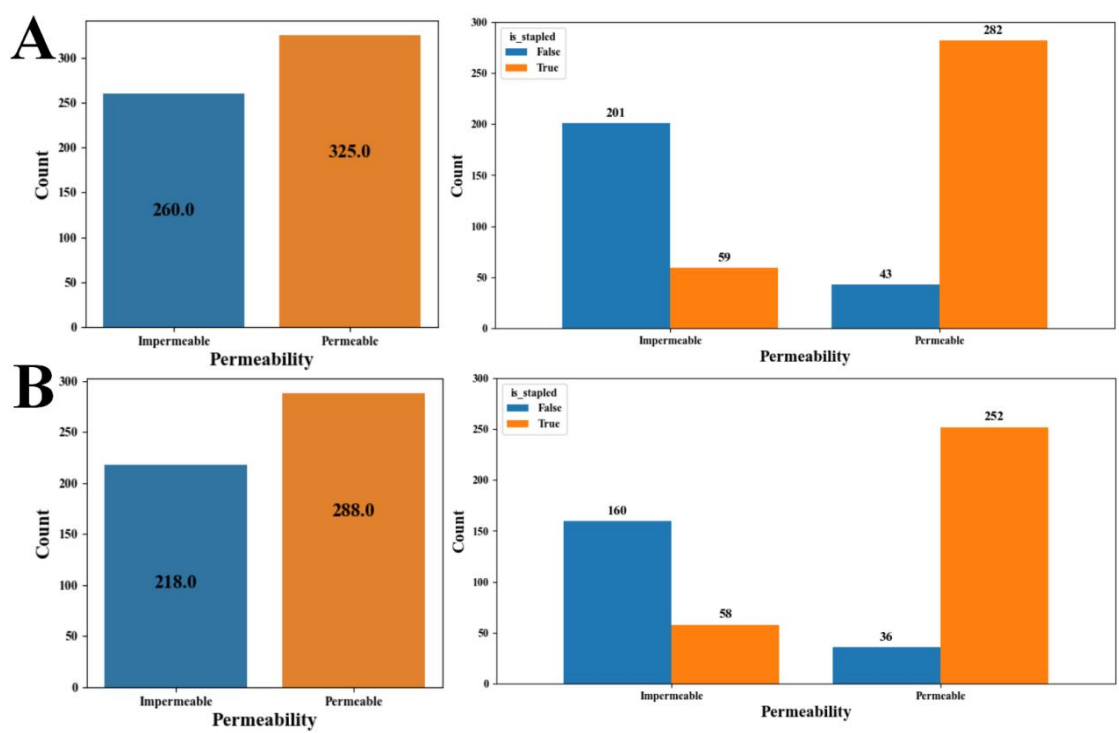

**Figure 2** (A) Permeability distributions of hydrocarbon-stapled peptides and non-hydrocarbon-stapled peptides in the original dataset. (B) Permeability distributions of hydrocarbon-stapled peptides and non-hydrocarbon-stapled peptides in the filtered dataset.

Out of the 21 calculated molecular descriptors, we selected 15 molecular descriptors to train the classification model, as illustrated in Figure 3. The reason why we selected these descriptors can be concluded as below. Descriptors like solvent-accessible surface area (sasa), accessible surface area (asa) and polar surface area (psa) represent the surface area of a biomolecule that is accessible to a solvent. Notably, psa is distinct as it specifically highlights the surface area occupied by polar atoms. Its strong correlation with passive molecular transport across membranes makes it a vital descriptor in studying drug transport properties, such as intestinal absorption and blood-brain barrier penetration[27]. Therefore, psa is chosen as the basis for constructing ML models. We select the hydrophobic index calculated from the structure, rather than from the sequence, as it better reflect the peptide's actual state. Furthermore, because of the linear correlation between the hydrophobic index and the lyticity index, the hydrophobic index is chosen as the basis for constructing the ML models. Similarly, since charge density is calculated from charge, we select the charge as a feature for building the ML models.

The overall distribution of 15 molecular descriptors for both permeable and impermeable peptides were generated using the Sta-ML module and displayed in Figure 3. The means of the 15 molecular descriptors for permeable and impermeable peptides are very close, with the means of charge for both permeable and impermeable peptides overlapping. It is important to note that a majority of the peptides exhibit a sheet percent of zero. Two potential reasons can be inferred: (1) Intramolecular cross-links in peptides can stabilize various secondary structures, with different cyclization sites stabilizing different secondary structures[28]. Cross-links between the side chains at positions i and i + 4 or/and i + 7, as well as hydrogen bond surrogate cross-links, tend to stabilize α-helices. Conversely, side chain-to-side chain, head-to-tail, and side chain-to-tail cyclization are more likely to stabilize structures such as turns, loops and β structures (β-sheets and β-strands). (2) Without a specific design or stabilization mechanism, linear peptides tend to form α-helical or randomly coiled structures. Considering the above reasons and the composition of our dataset, which comprises hydrocarbon-stapled peptides with side chain cross-links between i and i + 4 or/and i + 7, and linear peptides in a 310:196 ratio, it becomes clear why the sheet percent is zero in most peptides.

Principally relying on hydrogen bonding patterns and certain geometric constraints, the Define Secondary Structure of Proteins (DSSP) method assigns each residue to one of eight possible states[29]. Usually, secondary structure prediction methods simplify these eight states into just three main categories: α-helix, β-sheet, and loops. Since every residue in a peptide can be categorized as helix, sheet, or loop, there are inherent correlations among helix percent, loop percent, and sheet percent. Therefore, we only selected the helix percent and loop percent for further model construction.

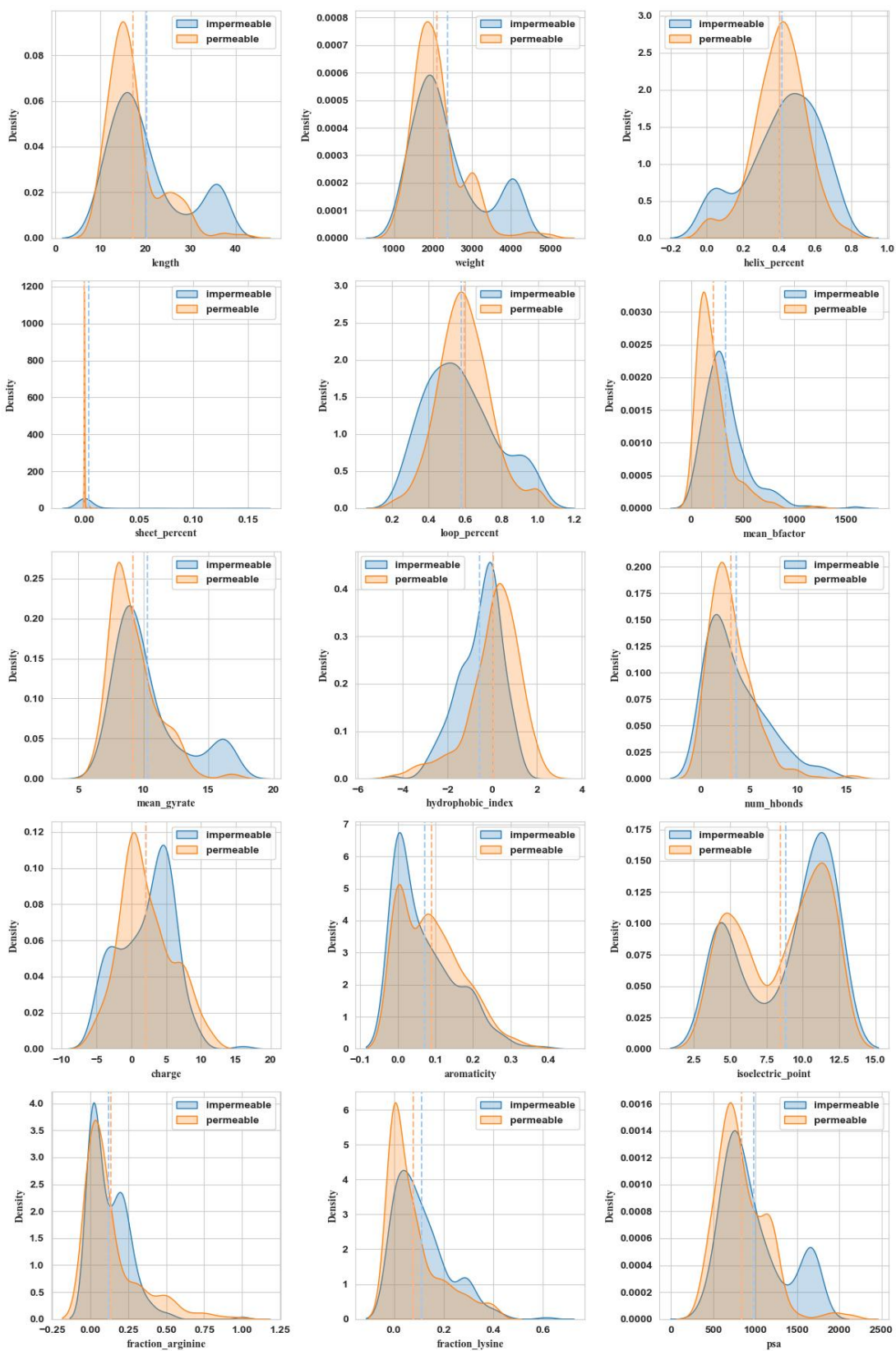

**Figure 3** The density curves of 15 molecular descriptors for permeable and impermeable peptides in the filtered dataset. The vertical dashed lines are the means of each group.

## 2.2 ML modeling results

The LightGBM classifier was constructed based on 14 features calculated by StaPep for distinguishing permeable and impermeable peptides. As shown in Figure 4A, the performance of this model is well on the test set with Accuracy, Recall, F1 and ROC-AUC all above 0.85. The ranking of molecular descriptors in terms of their importance was assessed using the 'feature_importances_' attribute of the LightGBM classifier. This analysis resulted in a ranking of the 14 molecular descriptors based on their significance in classifying permeable and impermeable peptides, as shown in Figure 4B. Although the mean values of the 15 features for both permeable and impermeable peptides are very close, each feature distributions exhibit variations. For instance, the differences in skewness and kurtosis of the hydrophobic index show two distinctive distributions between permeable and impermeable peptides. It is understandable that the difference may not be significant for a single feature alone (as that would imply the feature itself can effectively distinguish permeable and impermeable peptides). However, each feature contributes to the overall classification.

Figure 4B shows that the hydrophobic index, molecular weight, and psa are the top three features in terms of importance scores. This suggests that these features are potentially more critical than the other 11 in determining the membrane penetration property of peptides. Molecules with a higher lipophilicity tend to have a greater affinity for lipid-rich environments, such as cell membranes, which can impact their permeability and bioavailability[30]. Lipophilicity is a property that represents the affinity or tendency of a molecule to dissolve in or interact with lipids or non-polar solvents. It is composed of two components: hydrophobicity and polarity[31]. Hydrophobicity reflects the molecule's preference for non-polar environments or its aversion to water (hydrophobic nature), whereas polarity indicates the presence of polar or charged groups within a molecule that can interact with water or other polar solvents. Molecules with higher hydrophobic index are generally more soluble in non-polar solvents. Polar molecules have a degree of charge separation and are attracted to water molecules (hydrophilic nature). The balance between a molecule's hydrophobic and polar characteristics defines its overall lipophilicity. The hydrophobic index distribution in Figure 3 also indicates that permeable peptides have a higher mean hydrophobic index than impermeable ones. This data supports the inference that the hydrophobic index is a significant factor in the permeability of hydrocarbon-stapled peptides. For example, cyclic peptides primarily composed of hydrophobic residues can directly penetrate the plasma membrane through passive diffusion[30].

The permeability of a molecule generally has an inverse relationship with its molecular weight, implying that smaller molecules tend to have higher membrane diffusion rates[30]. An increase in molecular weight typically reduces a peptide's ability to cross epithelial cell membranes and enter systemic circulation[32]. Backbone cyclization reduces the peptide's size and enhances its conformational rigidity compared to linear counterparts, thereby increasing permeability. The Lokey group reported a significant decrease in permeability of lipophilic cyclic peptides as their size increased from 1000 to approximately 1400 Å$^3$ [33]. Furthermore, the weight distribution in Figure 3 also indicates that permeable peptides tend to have a lower mean molecular weight than impermeable peptides. The evidence supports the inference that the weight is a key factor affecting the permeability of hydrocarbon-stapled peptides.

psa measures the surface area of a peptide that is exposed to polar solvents, and is a highly reliable predictor for oral bioavailability and cellular permeability[30]. Molecules with a psa value exceeding 130 Å is highly associated with poor oral bioavailability, even if they adhere to other Ro5 properties[30]. Peptides with a lower psa often exhibit better membrane permeability, perhaps because they can more easily cross the non-polar, hydrophobic region of the cell membrane. Peptide cyclization can promote the formation of intramolecular hydrogen bonds, consequently reducing psa and enhancing passive cellular permeability[34,35]. The psa distribution in Figure 3 also indicates that permeable peptides have a lower mean psa than impermeable peptides. The evidence presented above supports the inference that psa plays a significant role in determining the permeability of hydrocarbon-stapled peptides.

The results discussed above provide additional support for the rationality and validity of the developed models. While these three features are pivotal, the remaining 11 features also play important roles in distinguishing between permeable and impermeable peptides. For example, features such as charge and the number of hydrogen bonds are also reported to be important parameters related to the permeability[36]. These molecular features can provide guidance for the strategic design of hydrocarbon-stapled peptides with optimized permeability.

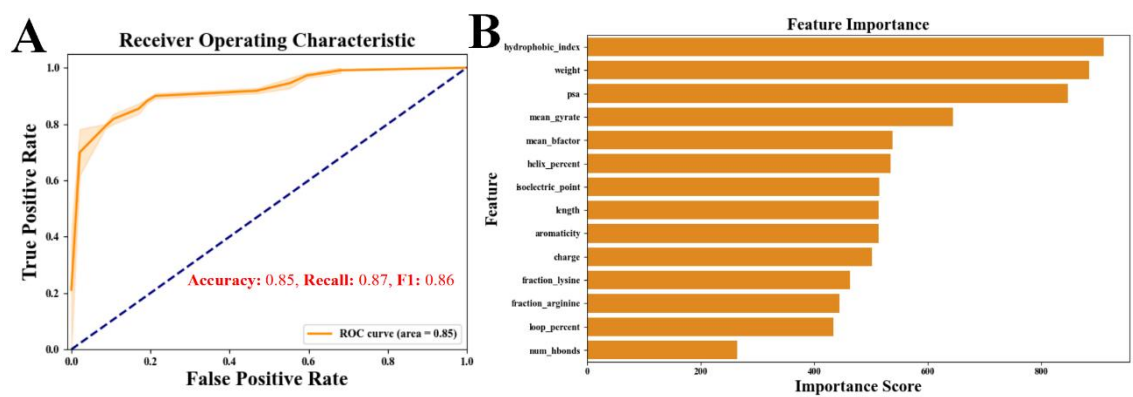

**Figure 4** (A) The Accuracy, Recall, F1 and ROC-AUC of the LightGBM model on 5-fold cross-validation. (B) The feature importance.

## 3 Discussion and Conclusion

Hydrocarbon-stapled peptides are emerging as potent regulators of protein-protein interactions, characterized by their high affinity and specificity for targeting intracellular molecules. Researchers can experimentally evaluate certain features of these peptides, such as hydrophobicity via high-performance liquid chromatography, helical content through circular dichroism, cellular uptake using confocal laser scanning microscopy, and biological activity through both in vitro and in vivo methods. However, there is a lack of tools calculating their structural and physiochemical descriptors, which limits the development of ML models for predicting their related biological properties, such as the cellular permeability. StaPep offers the first molecular feature toolkit designed specifically for hydrocarbon-stapled peptides. It is equipped with standard algorithms for feature calculation, ML modeling, and data visualization. We have integrated these essential functionalities into a single toolset, enabling users to easily create complex and flexible workflows. While StaPep offers the mentioned functionalities, we recognize some limitations that can be addressed in the future. This provides opportunities for new researchers and for our team to further refine and expand the toolkit's functionalities.

In the current version of StaPep, feature calculation is limited to hydrocarbon-stapled peptides containing 2 non-standard amino acids (norleucine and 2-aminoisobutyric acid) and 6 nonnatural anchoring residues (S3, S5, S8, R3, R5 and R8). In the future, we plan to expand StaPep's capabilities by integrating additional algorithms for calculating an even broader range of features for hydrocarbon-stapled peptides. Moreover, we aim to broaden the tool's scope to encompass all types of stapled peptides, and potentially extend its utility to cyclic peptides as well. StaPep is expected to become a comprehensive and powerful toolkit for investigating the scientific issue of stapled peptide in all aspects. StaPep is available as an open-source toolkit, with its source code accessible at https://github.com/dahuilangda/stapep_package.

## 4 Methods

### 4.1 Implementation

The StaPep developed in Python 3 within the Anaconda environment, using third-party modules and functions to run cheminformatics analysis. These tools can be installed through the available source codes, or using Conda virtual environments. In its current release, StaPep has six customized modules: (1) Sta-Structure, (2) Sta-feature, (3) Sta-MD, (4) Sta-3Dfeature, (5) Benchmark datasets and (6) Sta-ML, as detailed in Table 1. A guide to run these modules is available in Github (https://github.com/dahuilangda/stapep_package).

Table 1 Description of StaPep.

| No | Modules | Functions | Major libraries |
|---|---|---|---|
| 1 | Sta-Structure | Generate structure from the sequence of hydrocarbon- | pytraj*, ESMFold, Modeller |
| | | stapled peptide | |
| 2 | Sta-feature | Generate features from the sequence of hydrocarbon-stapled peptide | networkx*, matplotlib*, rdkit*, biopython* |
| 3 | Sta-MD | Perform molecular dynamics simulation of hydrocarbon-stapled peptide | openmm*, ambertools |
| 4 | Sta-3Dfeature | Generate features from the 3D structure of hydrocarbon-stapled peptide | parmed*, pytraj* |
| 5 | Sta-ML | Build benchmark datasets, apply filters and construct features-based ML models | pandas*, numpy*, CD-HIT*, lightgbm*, scikit-learn* |

*Python libraries used in data science

## 4.2 Feature extraction

Feature extraction is the core function of StaPep, utilizing various logical and mathematical formulas to convert the chemical information of hydrocarbon-stapled peptides into useful numbers. In the realms of chemoinformatics and bioinformatics, molecular features are crucial, particularly in ligand-based drug design, and represent the initial and most critical step in data analysis. The module "Sta-feature" can calculate 10 molecular features from the sequence of hydrocarbon-stapled peptides (Table 2). The module "Sta-Structure" has been developed to generate the 3D structure of hydrocarbon-stapled peptides. The non-standard amino acids in the peptide sequence are replaced with alanine (ALA). The 3D structure of the peptide sequence is first generated by ESMFold[37]. After obtaining the predicted 3D structure (PDB file), we edited the PDB file to replace the three-letter code of ALA with that of the non-natural amino acid, and adjust the atom names and coordinates accordingly. The basic simulation preparatory tool called tleap in AmberTools (https://ambermd.org/AmberTools.php) is utilized to build a protein system for the hydrocarbon stapled peptide in solvent. This process encompasses several critical steps: adding missing hydrogen atoms, defining force field parameters for the non-natural amino acids, solvating the protein, and adding salt ions, etc. The module "Sta-MD" can perform molecular dynamics simulation using OpenMM[38] to generate a stable 3D structure of the hydrocarbon stapled peptide. The module "Sta-3Dfeature" can calculate another 11 molecular features for hydrocarbon stapled peptide from its predicted 3D structure (Table 2). Overall, 21 molecular features of hydrocarbon-stapled peptides could be easily calculated by the StaPep based on sequence and structure (Table 2), which covers wide applications in various cheminformatics and bioinformatics projects.

**Table 2** List of various descriptors calculated by StaPep

| No | Method | Details |
|---|---|---|
| 1 | Name: seq_length<br>Type: for sequence | The length of the amino acid sequence (without N-terminus and C-terminus). |
| 2 | Name: weight<br>Type: for sequence | Molecular weight. |
| 3 | Name: hydrophobicity_index<br>Type: for sequence | The hydrophobicity index is calculated using the Kyte-Doolittle method[39] as below:<br>$$HI = \frac{1}{N}\sum_{i=1}^{n}(H_i)$$<br>where $N$ is the length of the amino acid sequence, and $H_i$ is the hydropathy index of the $i$-th amino acid (table S1). |
| 4 | Name: calc_charge<br>Type: for sequence | The net charge (Charge) of a peptide at a given pH can be calculated using the following formula:<br>$$Charge = Round\left(\sum_{aa \in pos\_pK} aa\_content[aa] \cdot \frac{10^{(pK-pH)}}{10^{(pK-pH)}+1} - \sum_{aa \in neg\_pK} aa\_content[aa] \cdot \frac{10^{(pH-pK)}}{10^{(pH-pK)}+1}\right)$$<br>where pos_pK refers to the pK values of arginine (R), lysine (K), histidine (H) and N-terminus. And neg_pK refers to the pK values of aspartate (D), glutamate (E), C-terminus, tyrosine (Y) and cysteine (C)[15]. aa_content is a dictionary containing the number of each type of amino acid in the sequence, and aa is the name of the amino acid, pK is the pK$_a$ value of the amino acid (table S2), pH is the pH value at which the charge is being calculated, and the default pH is 7.0. |
| 5 | Name: calc_charge_density<br>Type: for sequence | The charge density of a peptide at a given pH, such as pH=7.0, can be calculated using the following formula:<br>$$Charge = Round\left(\frac{Charge}{Weight}\right)$$<br>Where charge represents the net charge of the peptide at the given pH, weight represents the molecular weight of the peptide. |

| 6 | Name: aromaticity<br>Type: for sequence | The Aromaticity of a peptide can be calculated using the following formula:<br>$$Aromaticity = \frac{A}{N}$$<br>Where $A$ is the number of aromatic amino acids in the peptide, the aromatic amino acids are phenylalanine (F), tryptophan (W), and Y. $N$ is the length of the amino acid sequence. |
|---|---|---|
| 7 | Name: fraction_arginine<br>Type: for sequence | The fraction of arginine residues ($F_{arg}$) within a peptide can be calculated using the following formula:<br>$$F_{arg} = \frac{N_{arg}}{N}$$<br>Where $N_{arg}$ is the number of arginine residues in the peptide, $N$ is the length of the amino acid sequence. |
| 8 | Name: fraction_lysine<br>Type: for sequence | The fraction of lysine residues ($F_{lys}$) within a peptide can be calculated using the following formula:<br>$$F_{lys} = \frac{N_{lys}}{N}$$<br>Where $N_{lys}$ is the number of lysine residues in the peptide, $N$ is the length of the amino acid sequence. |
| 9 | Name: isoelectric_point<br>Type: for sequence | The DTASelect algorithm[40] computes the isoelectric point ($pI$) of a peptide sequence using a multi-step approach. First, the function determines the percentage of negative ions for the C-terminus, N-terminus, K, R, H, aspartic Acid (D), glutamic Acid (E), C and Y. Each function generates a concentration ratio (CR). For positive groups,<br>$$CR = 10^{pK-pH}$$<br>Negative ions reverse this order:<br>$$CR = 10^{pH-pK}$$<br>The algorithm assesses charged amino acid groups and proposes pH values for charge calculation. The charge ($Z$) is determined as the sum of fractional charges of these groups.<br>$$Z = \text{Nterm} + \text{Cterm} + \alpha \cdot \frac{CR_K}{CR_K + 1} + \beta \cdot \frac{CR_R}{CR_R + 1} + \gamma \cdot \frac{CR_H}{CR_H + 1} + \delta \cdot \frac{CR_D}{CR_D + 1} + \varepsilon \cdot \frac{CR_E}{CR_E + 1} + \zeta \cdot \frac{CR_C}{CR_C + 1} + \eta \cdot \frac{CR_Y}{CR_Y + 1}$$<br>Where $\text{Nterm}$ and $\text{Cterm}$ represent N-terminus and C-terminus charges, α, β, γ, δ, ε, ζ and η represent the number of residue K, R, H, D, E, C and Y, respectively. $CR_K, CR_R, CR_H, CR_D, CR_E, CR_C$ and $CR_Y$ are CRs for specific residues.<br>The $pI$ is then calculated as the pH at which the net charge is zero:<br>$$pI = \text{pH value at which } Z = 0$$<br>The algorithm uses a simulated isoelectric focusing approach for efficient pH value proposals. The isoelectric point is determined when the charge $Z$ is close to zero. The method considers factors such as neighboring residues and modifications. |
| 10 | Name: lyticity index<br>Type: for sequence | The lyticity index is a valuable tool for discovering antimicrobial peptides. We calculated the index based on the formula provided by Mouratada et al.[41] as below: |

$$LI = \sum_{i=1}^{N}(H_i + H_{i+4}) + \sum_{i=1}^{N-1}(H_i + H_{i+3})$$

where $N$ is the length of the amino acid sequence, and $H_i$ is the hydrophobicity value of the the $i$-th amino acid (table S2).

| 11 | Name: calc_helix_percent<br>Type: for structure | The percentage of helix (PH) in the secondary structure of a peptide throughout a molecular dynamic's trajectory is determined using the DSSP method (https://biopython.org/docs/1.75/api/Bio.PDB.DSSP.html) with the following formula:<br>$$PH = \frac{\sum_{i=1}^{n} N_{helix}}{n \times N} \times 100\%$$<br>where n is the number of frames in the molecular dynamics trajectory, $N_{helix}$ is the count of the helix residues in the frame $i$, and $N$ is the length of the amino acid sequence. |
|---|---|---|
| 12 | Name:<br>calc_extend_percent<br>Type: for structure | The percentage of sheet (PS) in the secondary structure of a peptide throughout a molecular dynamic's trajectory is determined using the DSSP method (https://biopython.org/docs/latest/api/Bio.PDB.DSSP.html) with the following formula:<br>$$Ps = \frac{\sum_{i=1}^{n} N_{sheet}}{n \times N} \times 100\%$$<br>where n is the number of frames in the molecular dynamics trajectory, $N_{sheet}$ is the count of the sheet residues in the frame $i$, and $N$ is the length of the amino acid sequence. |
| 13 | Name: calc_loop_percent<br>Type: for structure | The percentage of loop (PL) in the secondary structure of a peptide throughout a molecular dynamic's trajectory is determined using the DSSP method (https://biopython.org/docs/1.75/api/Bio.PDB.DSSP.html) with the following formula:<br>$$Ps = \frac{\sum_{i=1}^{n} N_{loop}}{n \times N} \times 100\%$$<br>where n is the number of frames in the molecular dynamics trajectory, $N_{loop}$ is the count of the loop residues in the frame $i$, and $N$ is the length of the amino acid sequence. |
| 14 | Name:<br>calc_mean_bfactor<br>Type: for structure | The Mean B-factor of $C\alpha$ atoms ($B_{C\alpha}$) is the average B-factor of all $C\alpha$ atoms in a peptide. The ($B_{C\alpha}$) is calculated using the following formula:<br>$$B_{C\alpha} = \frac{\sum_{i=1}^{n} \sum_{j=1}^{N} B_{C\alpha}^{i,j}}{n \times N}$$<br>Where $n$ is the number of frames in the molecular dynamic's trajectory, $N$ is the length of the amino acid sequence, ($B_{C\alpha}^{i,j}$) is the B-factor of the α atom of residue ($j$) in frame ($i$). |
| 15 | Name: calc_mean_gyrate<br>Type: for structure | The Mean Gyration Radius ($R_{\text{gyration}}$) is the average gyration radius of a peptide throughout a molecular dynamic's trajectory, which measures the compactness of a stapled peptide structure and is calculated using the following formula:<br>$$R_{\text{gyration}} = \frac{\sum_{i=1}^{n} R_{\text{gyration}}^{i}}{N}$$<br>Where $n$ is the number of frames in the molecular dynamic's trajectory, $N$ is |

| | | the length of the amino acid sequence, $R^i_{\text{gyration}}$ is the gyration radius of the peptide in frame $i$. |
|---|---|---|
| 16 | Name: calc_hydrophobic_index Type: for structure | The fixed hydrophobicity index (FHI) was calculated using the Kyte-Doolittle method[39] as below: $$FHI = \frac{1}{N}\sum_{i=1}^{n}\left(H_i \cdot \delta(A_i - MinA)\right)$$ where $N$ is the length of the amino acid sequence, $H_i$ is the hydropathy index of the $i$-th amino acid (table S1), $\delta$ is a Dirac delta function that takes the value of 1 when its argument is zero and 0 otherwise, and $A_i$ is the relative accessible surface area value (A-value) for the $i$-th amino acid calculated by Biopython's DSSP integration based on 3D structures[42]. $MinA$ is the threshold for filtering amino acids based on their A-value, only amino acids with an A-value greater than this threshold can be used in above formula, and the default value is 0.2 (20% of the maximum A-value). |
| 17 | Name: calc_mean_molsurf Type: for structure | Calculating the mean sasa within a dynamic's trajectory for a stapled peptide. |
| 18 | Name: calc_asa_for_average_structure Type: for structure | Calculating the mean asa using the optimized PDB structure of a hyrdocarbon-stapled peptide. |
| 19 | Name: calc_psa Type: for structure | Calculating 3D psa, which is a sum of surfaces of polar atoms (usually oxygens, nitrogens and attached hydrogens) in a hyrdocarbon-stapled peptide. |
| 20 | Name: calc_n_hbonds Type: for structure | Calculating total number of hydrogen bonds. |
| 21 | Name: extract_2d_structure Type: for structure | Obtaining SMILES from the 3D structure of a stapled peptide. |

## 4.3 A proposal for building ML models

### 4.3.1 Construction of benchmark datasets

A high-quality benchmark dataset is essential for developing an accurate and practical classification predictor. Data of hydrocarbon-stapled peptides and linear peptides with experimental cell permeability parameters were manually collected from references downloaded from the PubMed (https://pubmed.ncbi.nlm.nih.gov/). Accessed 1 April 2022. To ensure a comprehensive and targeted search, we used the keywords 'stapled peptide permeability' in PubMed, focusing on studies specifically reporting the cell permeability of hydrocarbon-stapled peptides. Finally, 341 hydrocarbon-stapled peptides and 244 linear peptides with sequence information and experimental membrane permeability data were collected (Figure 2), which can be downloaded from examples file in the code respository at https://github.com/dahuilangda/stapep_code/tree/master/example/datasets. To refine the dataset, duplicate peptides were removed and the peptide lengths were limited to between 10 and 50 amino acids. Subsequently, the modules of StaPep (Table 1) were applied automatically to extract the features of these peptides.

**4.3.2 The settings of ML experiments**

StaPep provides a set of standard utilities for ML modeling within its "Sta-ML" module. Within this module, we have included a built-in case study: a classification task for stapled peptides, utilizing LightGBM as the predictive model. To optimize the model's performance, a GridSearchCV with 5-fold cross-validation is employed on training set, allowing for the systematic exploration of optimal hyperparameter settings for the LightGBM model. Subsequently, the test set is used to evaluate the performance of the LightGBM models. The classification model's performance is measured using four key evaluation metrics: Accuracy, F1 score, AUC and Recall[43]. Additionally, feature importance is calculated based on the best estimator derived from the LightGBM model. The ".feature_importances_" attribute in LightGBM provides a measure of the relative importance of each feature in the gradient boosting tree model, helping identify which features have the most impact on the model's predictions. All the plots were created using the Python matplotlib package.

## Data availability

The hydrocarbon stapled-peptide dataset is available in Github (https://github.com/dahuilangda/stapep_package/tree/master/stapep/example/datasets).

## Code availability

The source code for StaPep implementation and data analysis are available in Github (https://github.com/dahuilangda/stapep_package).

## Author contributions

Z.W., X.L., S.C. and G.X. developed the concepts for the manuscript and proposed the method. Z.W., J.W., M.Z., C.G., B.Z., W.Z. and H.W. collected the data. Z.W., S.C. and X.L. designed the analyses and applications and discussed the results. Z.W. and S.C. conducted the analyses. X.L. and G.X. helped interpret the results of the real data analyses. Z.W., S.C. and X.L. prepared the manuscript and contributed to editing the paper.

## Competing interests

The authors declare no competing interests

## ABBREVIATIONS USED

AUC=area under the curve

ML=machine learning

psa=polar surface area

sasa=solvent-accessible surface area

asa=accessible surface area

FHI=fixed hydrophobicity index

$R_{gyration}$=Mean Gyration Radius

$B_{C\alpha}$=Mean B-factor of C$\alpha$ atoms

PL=The percentage of loop

PS=The percentage of sheet

PH=The percentage of helix

pI=isoelectric point

H=histidine

D=aspartic Acid

E=glutamic Acid

C=cysteine

CR=concentration ratio

Z=The charge

$F_{lys}$=The fraction of lysine residues

$F_{arg}$=The fraction of arginine residues

F=phenylalanine

W=tryptophan

Y=tyrosine

R=arginine

K=lysine

D=aspartate

E=glutamate

ALA=alanine

DSSP=Define Secondary Structure of Proteins

# Supplementary

## 1. Calculating the hydropathy index of anchors and non-standard amino acids

Linear regression follows the linear mathematical model for determining the value of one dependent variable from value of one given independent variable. scikit-learn was applied to build the linear regression model. The hydropathy indexes of natural amino acids were collected from a reference[1]. While hydrophobicity values in table S3 were define as the independent variable, hydropathy indexes in table S1 were define as the dependent variable. The "hydropathy indexes" of X (S5), S3, S8, R3, R5, R8, Aib and B in table S1 were predicted based on related "hydrophobicity values" by this linear regression model.

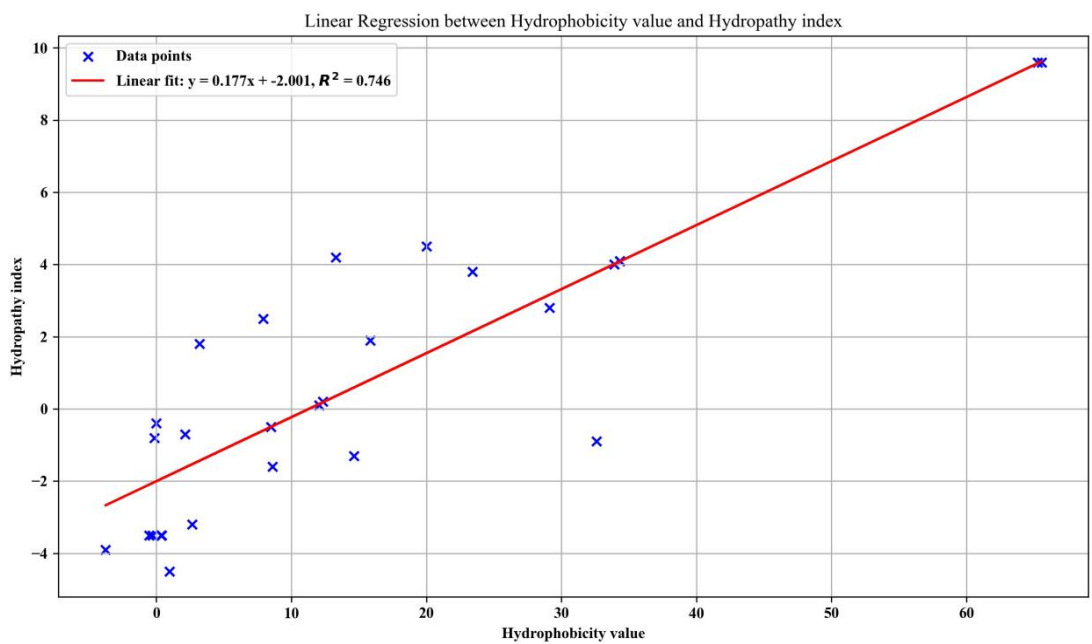

**Figure S1** The linear regression model with $R^2$ equals to 0.746.

**Table S1** The hydropathy indexes of amino acids and anchors

| Amino Acid or Anchor | Hydropathy index | Amino Acid or Anchor | Hydropathy index |
|---|---|---|---|
| A | 1.8 | P | -1.6 |
| R | -4.5 | S | -0.8 |
| N | -3.5 | T | -0.7 |
| D | -3.5 | W | -0.9 |
| C | 2.5 | Y | -1.3 |
| Q | -3.5 | V | 4.2 |
| E | -3.5 | X (S5) | 4.1 |
| G | -0.4 | S3 | 0.1 |
| H | -3.2 | S8 | 9.6 |
| I | 4.5 | R3 | 0.2 |
| L | 3.8 | R5 | 4.0 |
| K | -3.9 | R8 | 9.6 |
| M | 1.9 | Aib | -0.5 |
| F | 2.8 | B | 2.3 |

**Table S2** The p$K_a$ values of amino acids and anchors[2]

| Amino Acid or Anchor | p$K_a$ value |
|---|---|
| pos_pk | |
| R | 12.1 |
| K | 10.67 |
| H | 6.04 |
| N-terminus | 9.38 |
| neg_pk | |
| D | 3.71 |
| E | 4.15 |
| C-terminus | 2.15 |
| Y | 10.10 |
| C | 8.14 |

## 2. Testing the hydrophobicity values of amino acids and anchors

### 2.1 Material

All amino acids were obtained from GL Biotech. Rink amide MBHA resin (0.37 mmol/g) was purchased from Tianjin Nankai Hecheng Science &Technology Co. Ltd. All reagents and solvents such as dichloromethane (DCM), N, N-dimethylformamide (DMF) were purchased from Chemical Reagent Co. Ltd. or Titan Scientific Co. Ltd.

### 2.2 Peptide Synthesis and Purification

We employed solid-phase peptide synthesis (SPPS) to synthesize the linear peptide, with Rink amide MBHA resin serving as the solid-phase support. First, the Rink amide MBHA resin was swollen in dichloromethane at room temperature for 20 min, followed by the deprotection of the Fmoc group. The resin was subsequently rinsed with DMF and DCM. Then the coupling of Fmoc-amino acids were performed for 30 min. The deprotection, coupling, and washing steps were repeated until the last amino acid residues were linked to resin. Next, the peptide-bound resin was treated with acetic anhydride and pyridine (1:1) for 20 min at room temperature. For the stapled peptides, Grubbs' first-generation catalyst (10 mM) was used for cyclization in RCM reaction. After that, the crude peptides were cleaved from the resin using a cocktail reagent (88% TFA, 2% TIPS, 5% water, and 5% phenols) for 4 h. Finally, the peptides were crystallized with cold ether and subsequently analyzed and identified using HPLC and MS.

The analytical HPLC setup featured an analytical column (XBridge C18, 5 μm particle size, 150 mm × 4.6 mm) with a flow rate of 1.0 mL/min. Analytical injections were monitored at 214 nm at room temperature. Mobile phase A consisted of 0.1% TFA in water, while mobile phase B was composed of 0.1% TFA in acetonitrile. A gradient elution program was applied, transitioning from 5% to 65% of mobile phase B over 25 minutes (from 5 to 30 minutes). For peptide separation, a preparative column (YMC-Pack ODS-AQ C18, 10 μm particle size, 250 mm × 20 mm) was utilized with a flow rate of 20 mL/min. According to the polarity of each peptide, the separation method is not entirely the same. The mobile phase B was changed with a gradient from 30% to 90% over a period of 60 minutes.

## 2.3 HPLC Analysis of Synthetic Model Peptides

Mobile phase A consisted of a 10mM $NaH_2PO_4$ aqueous solution adjusted to pH 7 using NaOH. Mobile phase B was prepared by mixing 50% mobile phase A and 50% acetonitrile. Retention times were determined using an analytical column (Waters 'XBridge C18', 4.6 × 150 mm, 5 μm particle size) with a flow rate of 1.0 mL/min. The target peptides were monitored at 214 nm at room temperature, with a linear gradient of 2% to 62% mobile phase B over a duration of 115 minutes (from 5 to 120 minutes).

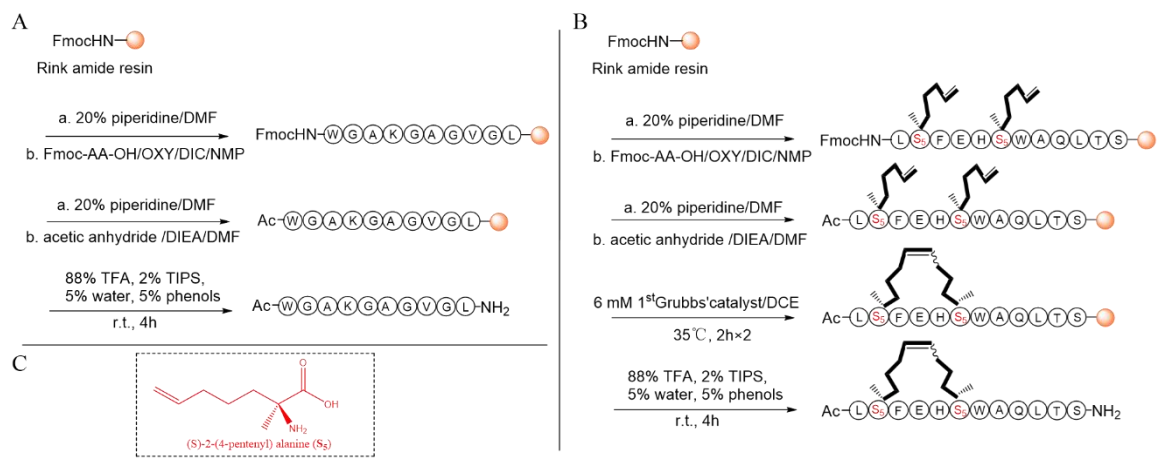

**Figure S2** (A) The synthetic route of the linear peptide. (B) The synthetic route of the stapled peptide. (C) Chemical structure of the (S)-2-(4-pentenyl) alanine (S5) that cross-linked by ring-closing metathesis.

**Table S3** The hydrophobicity values determined by RP-HPLC of amino acids and anchors

| Amino Acid Substitution | Sequence | Retention Time (RT, min) | Hydrophobicity values* |
|---|---|---|---|
| W | Ac-WGAKGAGVGL-NH$_2$ | 73.452 | 32.606 |
| F | Ac-FGAKGAGVGL-NH$_2$ | 69.967 | 29.127 |
| L | Ac-LGAKGAGVGL-NH$_2$ | 64.244 | 23.398 |
| I | Ac-IGAKGAGVGL-NH$_2$ | 60.860 | 20.014 |
| M | Ac-MGAKGAGVGL-NH$_2$ | 56.700 | 15.854 |
| Y | Ac-YGAKGAGVGL-NH$_2$ | 55.493 | 14.647 |
| V | Ac-VGAKGAGVGL-NH$_2$ | 54.131 | 13.285 |
| P | Ac-PGAKGAGVGL-NH$_2$ | 49.439 | 8.593 |
| C | Ac-CGAKGAGVGL-NH$_2$ | 48.752 | 7.906 |
| A | Ac-AGAKGAGVGL-NH$_2$ | 44.053 | 3.207 |
| E | Ac-EGAKGAGVGL-NH$_2$ | 41.232 | 0.392 |
| T | Ac-TGAKGAGVGL-NH$_2$ | 42.963 | 2.117 |
| D | Ac-DGAKGAGVGL-NH$_2$ | 40.572 | -0.274 |
| Q | Ac-QGAKGAGVGL-NH$_2$ | 41.184 | 0.338 |
| S | Ac-SGAKGAGVGL-NH$_2$ | 40.689 | -0.157 |
| N | Ac-NGAKGAGVGL-NH$_2$ | 40.323 | -0.523 |
| G | Ac-GGAKGAGVGL-NH$_2$ | 40.846 | 0 |
| H | Ac-HGAKGAGVGL-NH$_2$ | 43.499 | 2.653 |
| K | Ac-KGAKGAGVGL-NH$_2$ | 37.072 | -3.774 |
| R | Ac-RGAKGAGVGL-NH$_2$ | 41.827 | 0.981 |
| R$_3$ | Ac-R$_3$GAKGAGVGL-NH$_2$ | 53.166 | 12.32 |
| S$_3$ | Ac-S$_3$GAKGAGVGL-NH$_2$ | 52.896 | 12.05 |
| R$_5$ | Ac-R$_5$GAKGAGVGL-NH$_2$ | 74.748 | 33.902 |
| X (S$_5$) | Ac-S$_5$GAKGAGVGL-NH$_2$ | 75.166 | 34.32 |
| R$_8$ | Ac-R$_8$GAKGAGVGL-NH$_2$ | 106.118 | 65.272 |
| S$_8$ | Ac-S$_8$GAKGAGVGL-NH$_2$ | 106.401 | 65.555 |
| NLE | Ac-(NLE)GAKGAGVGL-NH$_2$ | 65.288 | 24.442 |
| Aib | Ac-Aib-GAKGAGVGL-NH$_2$ | 49.339 | 8.493 |

*hydrophobicity value = RT$_{amino\ acids}$-RT$_G$